\documentclass[aps,reprint,10pt,twocolumn,superscriptaddress]{revtex4}
\usepackage{amsmath}
\usepackage{amssymb}
\usepackage{bbm}
\usepackage{graphicx}
\usepackage[countmax]{subfloat}
\usepackage{framed}
\usepackage{color}
\usepackage{xcolor}
\usepackage{hyperref}
\usepackage{epstopdf}
\usepackage{dsfont}
\usepackage{amsthm}
\usepackage{wrapfig}
\usepackage{relsize}
\usepackage{bm}
\usepackage[latin1]{inputenc}
\usepackage{enumitem}
\usepackage{soul}
\usepackage{float}

\newcommand{\red}[1]{{\color{red}}}

\usepackage{pifont}

\newcommand{\ba}{\begin{eqnarray}}
\newcommand{\ea}{\end{eqnarray}}
\begin{document}

\title{Neural networks for detecting multimode Wigner-negativity} 

\author{Valeria Cimini}\email{valeria.cimini@uniroma3.it}
\affiliation{Dipartimento di Scienze, Universit\`{a} degli Studi Roma Tre, Via della Vasca Navale 84, 00146, Rome, Italy}
\author{Marco Barbieri}
\affiliation{Dipartimento di Scienze, Universit\`{a} degli Studi Roma Tre, Via della Vasca Navale 84, 00146, Rome, Italy}
\author{Nicolas Treps}
\affiliation{Laboratoire Kastler Brossel, Sorbonne Universit\'{e}, CNRS, ENS-PSL Research University, Coll\`{e}ge de France, 4 place Jussieu, F-75252 Paris, France}
\author{Mattia Walschaers}
\affiliation{Laboratoire Kastler Brossel, Sorbonne Universit\'{e}, CNRS, ENS-PSL Research University, Coll\`{e}ge de France, 4 place Jussieu, F-75252 Paris, France}
\author{Valentina Parigi}
\affiliation{Laboratoire Kastler Brossel, Sorbonne Universit\'{e}, CNRS, ENS-PSL Research University, Coll\`{e}ge de France, 4 place Jussieu, F-75252 Paris, France}
%\affiliation{Istituto Nazionale di Ottica - CNR, Largo Enrico Fermi 6, 50125, Florence, Italy}

\begin{abstract} 
The characterization of quantum features in large Hilbert spaces is a crucial requirement for testing quantum protocols. In the continuous variables encoding, quantum homodyne tomography requires an amount of measurements that increases exponentially with the number of involved modes, which practically makes the protocol intractable even with few modes. Here we introduce a new technique, based on a machine learning protocol with artificial Neural Networks, that allows to directly detect negativity of the Wigner function for multimode quantum states. We test the procedure on a whole class of numerically simulated multimode quantum states for which the Wigner function is known analytically. We demonstrate that the method is fast, accurate and more robust than conventional methods when limited amounts of data are available. Moreover the method is applied to an experimental multimode quantum state, for which an additional test of resilience to losses is carried out. 
 \end{abstract}

\maketitle

The ability to engineer large and scalable multi-party quantum states is at the core of future quantum technologies. In particular, large entangled structures are essential for measurement-based quantum computing protocols \cite{Briegel,Gu09}.  Multimode quantum optics represents a powerful platform for generating large entangled networks in the continuous variables (CV) regime.  Over the last decade the generation of up to thousands multimode entangled states has been experimentally demonstrated~\cite{Yokoyama2013,Roslund14,Chen14,Asavanant19,Larsen369} in optical parametric processes. These quantum states, which are characterized by Gaussian statistics,  are necessary but not sufficient alone to perform quantum computing protocols, as they can be efficiently simulated via classical resources.  The implementation of unconditional non-Gaussian operations is a much more demanding task, as it requires strong non-linear interactions.  In the perspective of investigating intermediate-scale systems in the near term, degaussification via heralded photon subtraction/addition operations has been demostrated ~\cite{Wenger04,Zavatta04,JonasNN06,Ourjoumtsev06,Parigi07,Sasaki08,Ourjoumtsev09,Bimbard10,Namekata10,JonasNN10,Gerrits10,Lee11,Asavanant:17,Biagi20, Lvovsky20}. If the generation of non-Gaussian multimode entangled states
is within reach of state-of-the-art experimental platforms, the full characterization of their quantum state remains a hard task.

In the CV picture Gaussian quantum states are completely characterized by the mean values of two conjugated quadratures per mode, plus their covariance matrix. Beyond Gaussian statistics, the complete quantum description of the optical system in terms of the density matrix or its equivalent Wigner representation, may be recovered via quantum homodyne tomography. In usual maximum likelihood (MaxLik)  procedures  \cite{Lvovsky04, Mogilevtsev07} good accuracy in the state reconstruction requires a large number of measurements, which scales exponentially with the number of involved optical modes. This implies that the setup should be stable until the whole set of measurements is taken, and that the algorithm for reconstructing the state becomes computationally too heavy.

A particular feature that we are interested to recover for non-Gaussian quantum states is the negativity of their Wigner representation, as it is accounted to be a pivotal quantum resource \cite{Mari12,Takagi18,Albarelli18}. This can be tested after its tomographic reconstruction via the usual MaxLik procedure,  but this is not a viable option in the multimode scenario.  Moreover, when interested in a particular quantum feature, like the Wigner negativity, and not in the complete knowledge of the Wigner function, it is worth finding a more direct approach to link the measurements and the specific property we are interested in.

In this letter we discuss an alternative approach, easily scalable with the number of optical modes of the system, that aims to specifically identify  the presence of Wigner-negativity, given a set of quadrature data.
Our method relies on machine learning algorithms \cite{RevModPhysML}, which have been demonstrated to be particularly powerful for characterization and optimization of quantum systems in different contexts \cite{Carleo602, PhysRevLett.114.200501, Dunjko18, PhysRevApplied.10.044033, Hentschel1, Rocchettoeaau1946, Yu2019, Melnikov1221, PhysRevX.8.031086, PhysRevX.9.011013, Gebhart}. In particular, artificial Neural Networks (NNs) already offer an alternative and efficient strategy to represent quantum many-body states, enabling to perform quantum state tomography for high dimensional states from a limited number of experimental data \cite{Torlai_qst,PhysRevLett.123.230504}.  Most of the protocols have been implemented in a Discrete Variables framework, one approach for quantum homodyne tomography has been proposed \cite{Fedorov}, and experimentally tested in the single-mode configuration. 
 Our algorithm allows, in a supervised learning approach, the discrimination between multimode optical states presenting negative or positive Wigner function and it is the first application of a machine learning  algorithm to CV multimode optical states.
 
Compared to standard quantum homodyne tomography protocols, out method is more robust when limited data are available. Furthermore, it allows to identify the Wigner negativity for states up to ten modes, a task too hard to accomplish with current MaxLik procedures. In this letter we first test the method with simulated data and then we apply it to classify the Wigner negativity of experimental quantum states.
 
Testing the negativity of the Wigner function, namely study if the function is either always positive or shows some negative regions, can be seen as a binary classification problem \cite{Murphy2012}. This is a very common use case of machine learning algorithms, which are daily used to solve tasks like email spam filtering, document categorization, speech, image and handwriting recognition. 
The use of NNs is in fact suitable in problems where the outcomes of the observed variables span in a large space. This is the case for the measurement of optical quadratures whose continuous values span, in principle, over an infinite phase space. Indeed, thanks to the network's ability to dynamically create complex prediction functions, there is no need for modelling \cite{PhysRevLett.123.230502, Macarone2019, Fossel18, PhysRevA.95.012335}.
Also, in standard tomographic methods a considerable number of data has to be collected for every new instance to be classified and the algorithm is run with no memory of previous samples. On the contrary in the NNs approach, after a first stage of training (learning) on simulated quadrature measurements of hundreds of different states, the network is able to give a fast classification of new samples. The performance of this new method is then compared to the Wigner-negativities that are found when applying the MaxLik protocol to the same quadrature data for few-mode states.

The success of the NN approach in classifying states with Wigner-negativity relies on our ability to generate training data, {\it i.e.} to simulate quadrature outcomes of various states. Therefore, we limit ourselves to the actual operations that are at hand in the experiments of interest: the ability to prepare an arbitrary $m$-mode Gaussian state, the ability to act with a non-Gaussian operation on these states through either photon addition or photon subtraction, and the occurrence of losses. To generate the quadrature data, we use the Wigner function that can be obtained analytically \cite{PhysRevLett.119.183601,PhysRevA.96.053835,PhysRevA.100.023828}.  We simulate squeezed states over {m=3}, 5 or 10 modes (with randomly chosen squeezing between $0 {\rm dB}$ and $8 {\rm dB}$ for each mode), we randomly choose whether or not to add or subtract a photon, and we add a randomly chosen percentage of losses and thermal noise (see Supplementary Material for details \cite{SuppMat}). We generate $4000$ of these states, with approximately the same number of positive and negative Wigner functions. For each we choose a random mode-basis in which we perform $k=1000$ repetitions of joint quadrature measurements. Each of them contains three detectors outcome per mode associated to three different phases, chosen randomly within three fixed phase intervals. This leads to a total of $3000 \times m$ quadrature measurements per state.

Because the training data are simulated starting from an analytical Wigner function, we know its minimal value $W_{\rm min}$. This value can be converted to a binary classifier: the target output $W_0$ is set to $W_0=0$ if $W_{min}\ge C$, while $W_0=1$ if $W_{min}< C$. The constant $C < 0$ represents a cut-off allowing to exclude limiting cases where the Wigner-negativity is too close to zero to be considered significant or to look specifically for highly negative Wigner functions. Thus, for every simulated measurement we add the label $W_0$ to highlight whether or not the data correspond to a non-positive Wigner function.

Feeding in the complete set of joint homodyne detection events would require a NN of thousands of nodes. Instead, in order to pass the information contained in the quadrature distribution to the input layer, we binned the data for each individual mode and choice of phase, and evaluate the occurrence frequencies in each bin, normalized to the total number of measurements. 
In practice, we process the data with $5$ bins relative to the $3$ different phase values for each mode, which corresponds to having an input layer with only $15\times m$ nodes. We feed the network with a matrix of all combined training data with $15\times m$ columns and as many rows as the different states that we use for the training. With this method of data processing, we do not require joint quadrature measurements. The correlations between the different modes are effectively integrated out, thus the method scales linearly with the number of modes rather than exponentially. This also makes the generation of training data less demanding.

To identify the Wigner-negativity of an $m$-mode optical state we use a feed-forward NN with three hidden layers, with respectively $30$, $20$ and $10$ nodes, all activated by a rectified linear unit (reLU) function.  This architecture of the NN  has been selected among the
different tested configurations as the one yielding the least means squared error on the validation set. The output layer consists in one node activated by sigmoid function. The tuning of all the hyperparameters of the network, namely all those parameters that are set before the learning process begins, is done using a grid search in order to determine the optimal values for a given model. The training is performed minimizing the loss function, that in our case is a simple mean squared error between the target value $W_0$ and the network output. The later, that corresponds to the probability of having a negative minimum value of the Wigner function, is then used to classify the state as negative if it exceeds the threshold $P_{th}=50\%.$

As customary in evaluating the NN performance, we use a cross-validation procedure, in which we split the data into two parts - the training set ($80\%$ of total data), and the validation set ($20\%$ of total data). The training set is used to train the model, while the validation set is used to evaluate the model's performance on a different sample of data and it is used to stop the network training as soon as the loss function evaluated for this set stops decreasing in a sufficiently long interval, in order to avoid overfitting. This is referred to as early stopping. 

\begin{figure}[h]
\begin{center}
{\includegraphics[width=\columnwidth]{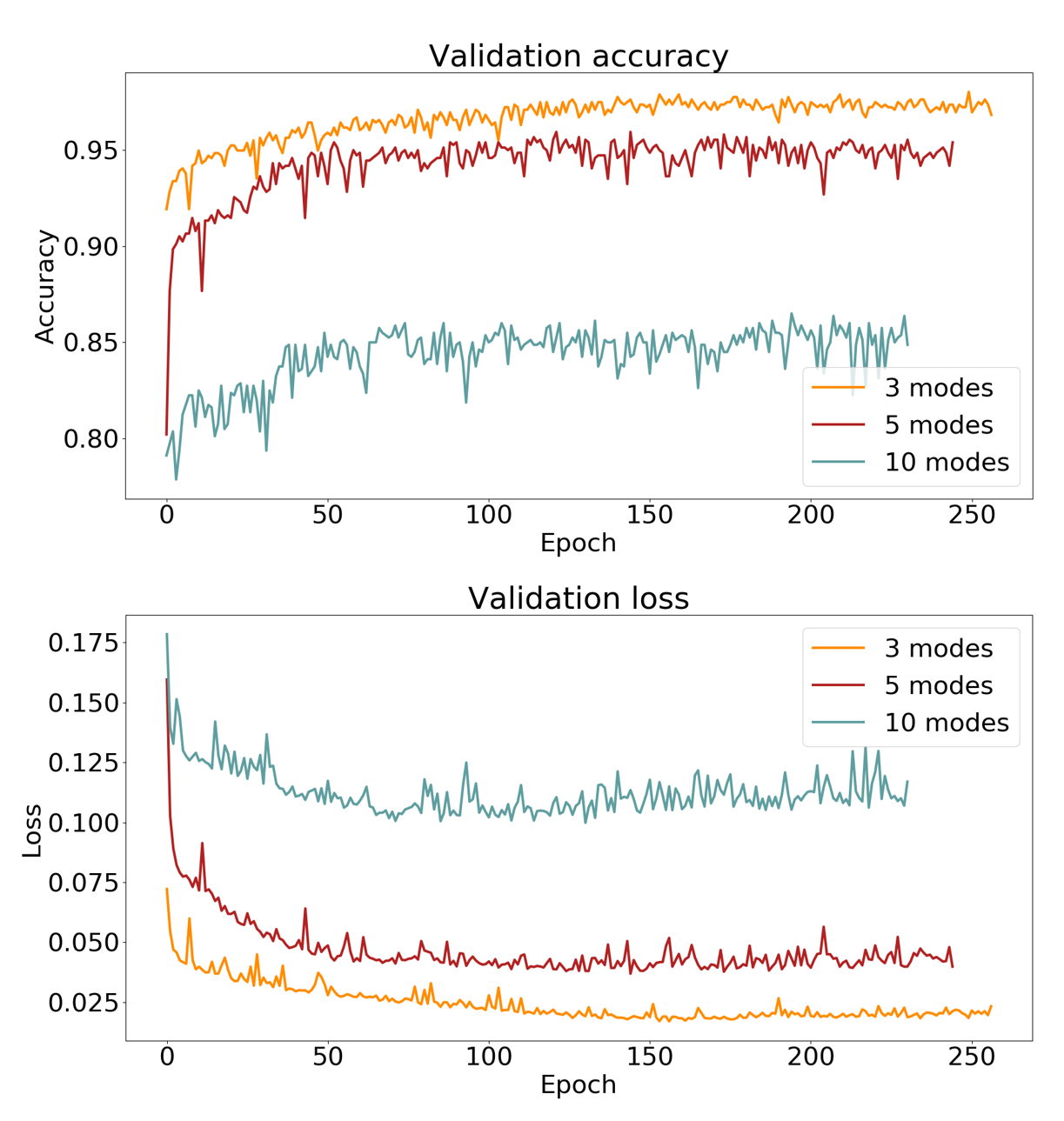}}
\caption{NN performances on the validation set. Top: accuracy of the model as a function of the epoch, i.e. each iteration of the training. Bottom: loss function as a function of the epoch. In both panels, the orange, red and cyan curves refer respectively to the performances on $m = 3, 5$ and $10$ modes states.}
\label{fig1}
\end{center}
\end{figure}

In Fig. \ref{fig1} we show how the model performs, after it has been trained, on the independent validation set. To this purpose, we evaluate the NN accuracy defined as the fraction of instances in the validation set which are correctly identified. The correct identification of states with a negative Wigner function happens with good accuracy for all the tested cases. The accuracy for the $m=3$ modes case is particularly high, exceeding $95\%$ of correct identifications, however even for the $m=10$ modes states, our method correctly identifies more than $85\%$ of the states in the validation set. We found that the discrimination is optimised using the cut-off $C=- 0.1/(2\pi)^m $, where $-1/(2\pi)^m$ is the maximal negativity attained by an $m$-mode state. We refer the reader to the Supplementary Material \cite{SuppMat} for further details on the algorithm and its appraisal.

We remark that even starting with the same batch of training data, the optimization of the NN will end up in slightly different configurations, since the weights of the connections are randomly initialized. In addition, there is an element of potential variability in using different training sets taken from the same class of states. However, we have checked that this level of accuracy is reproducible and independent on the choice of the training set.    

\begin{figure}[h]
\begin{center}
{\includegraphics[width=\columnwidth]{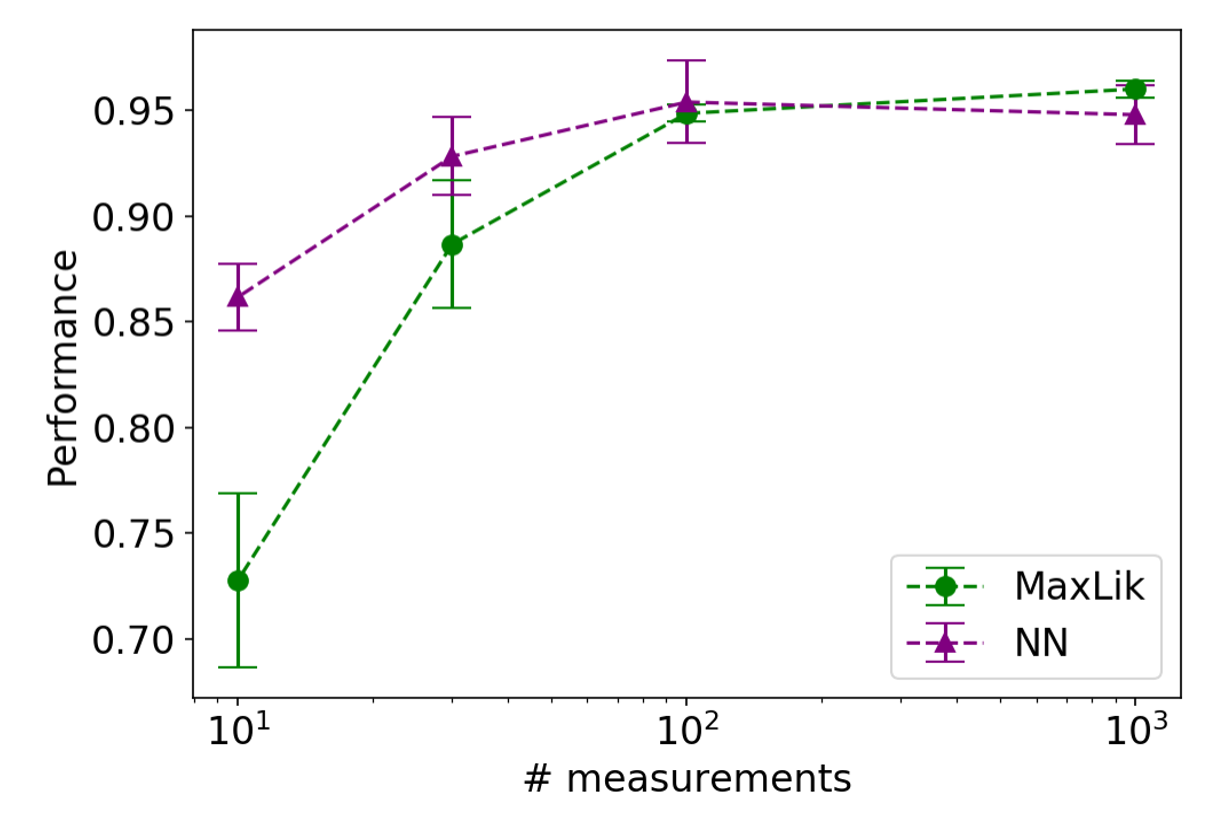}}
\caption{Comparison between MaxLik and NN performance as a function of the input data size. In both approaches, this is quantified as the fraction of states correctly classified. Error bars reflect the variability over $6$ repetitions. }
\label{fig2}
\end{center}
\end{figure}

We investigated the comparison between the NN performance in identifying the Wigner-negativity with the one of the standard state tomography, based on the MaxLik algorithm. This is carried out only for $m{=}3$, for which this procedure is computationally feasible. Even with a limited amount of data, this method will provide a density matrix $\rho_*$ for the state, represented in the Fock basis. It is reasonable to assume that $\rho_*$ already manifests Wigner-negativity long before the full tomography has converged. When we limit ourselves to photon-added or -subtracted Gaussian states that do not have any initial mean field, it can be proven that the Wigner function achieves its most negative value in the origin of phase space \cite{PhysRevLett.119.183601}. This value can be calculated \cite{PhysRevA.15.449} via the parity operator:
\begin{equation}\label{eq:maxlikneg}
    W_{min} = \sum_{n_1,n_2,n_3 = 0}^{N_{ph}} \frac{(-1)^{n_1+n_2+n_3}}{8\pi^3}\langle n_1,n_2,n_3 \rvert \rho_*  \lvert n_1,n_2,n_3 \rangle,
\end{equation}
where $\lvert n_1,n_2,n_3 \rangle$ denotes the state of the $3$-mode Fock basis, and $N_{ph}$ is the maximal photon number that is chosen as a trade-off between accuracy and computing time for the MaxLik procedure. Here, we set $N_{ph}=5$ and run $15$ iterations of the algorithm. In Fig.~\ref{fig2}, the performance of the NN approach is compared to that of MaxLik for the same number of measurements $k = 1 000, 100, 30, 10$, on $100$ different states. The uncertainties correspond to the variability observed over $6$ different batches. As expected, the MaxLik estimation improves its performance with the number of measurements and its variability improves as well. The adoption of NN results in a two-fold advantage when a reduced number of measurements is available: the value of the performance is more robust and also its variability remains more confined, with an improvement of a factor of about $40\%$ in standard deviation.
We propose that the same method can also be applied with similar success on actual experimental data. The key ability is then to know a sensible approximation of the main characteristics of the experimental setup such as losses, the maximum level of squeezing and the noise present in the state. By these, it becomes possible to train the NN using simulated data, and still obtain good accuracy in identifying the interesting feature of the experimental state. 

We test these ideas with an experimental single-photon subtracted $2-$mode entangled state whose generation has been reported in \cite{Ra19}. The initial state before subtraction shows the entanglement correlations of an EPR state \cite{PhysRevLett.68.3663} between quadratures of two given frequency-modes. The single photon subtraction takes place on one of the two-given frequency modes, so that, in case of high purity and low losses, the first mode is left in a vacuum state while the second mode is left in a state with a negative Wigner function (see Supplementary material \cite{SuppMat} for more details).
We train the network using 1000 simulated quadrature measurements, calibrated with the known imperfections of our experimental setup, namely thermal noise equal to $1.11$ (directly linked with the purity of the initial Gaussian state), a factor of $12\%$ of losses, and a maximum level of squeezing of $3$ dB. The data are binned as before in order to build the input to the NN. Using the actual experimental data we can compute the complete Wigner function and the density matrix of the state using a maximal photon number of only $N_{ph} = 3$, giving  $W_{min}=-0.03$ (see Supplementary Material \cite{SuppMat}). We now take 15000 experimental quadratures, arranged in $15\times2$ histograms, and we feed them to the NN. Our algorithm is able to detect the Negativity present in the experimental state. Since imperfections limit the minimum of the Wigner function, no cut-off, {i.e.} $C=0$, is used in this case.

The characteristics of the experiment can be obtained within a certain accuracy and precision. Misrepresentation of the experimental parameters may lead to a failure of the NN classification. On the other hand, these networks are known to be able to work reliably even in the presence of noise. This suggest that discrepancies between actual and simulation parameters can be tolerated. 

\begin{figure}[ht!]
\begin{center}
{\includegraphics[width=\columnwidth]{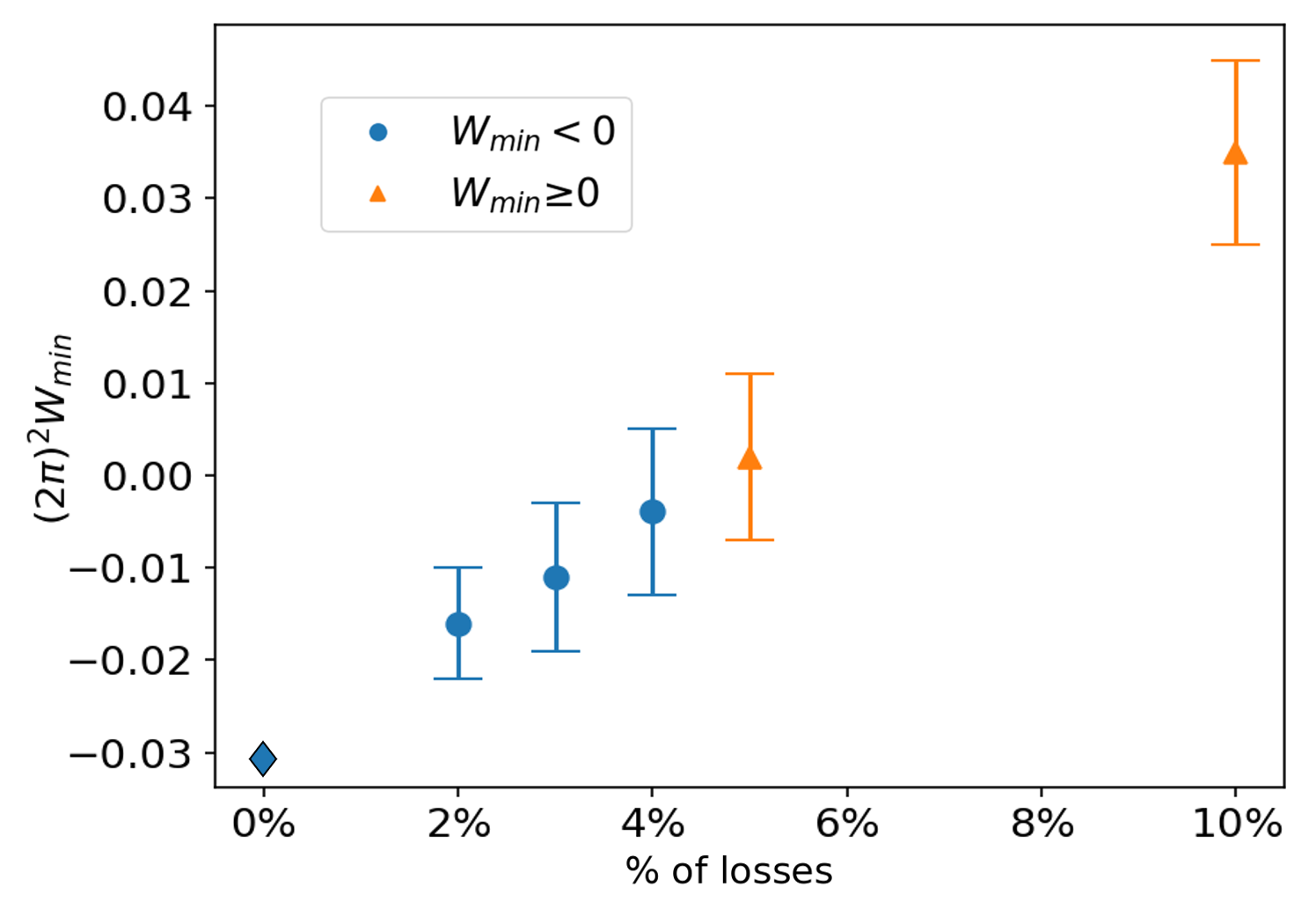}}\\
{\includegraphics[width=\columnwidth]{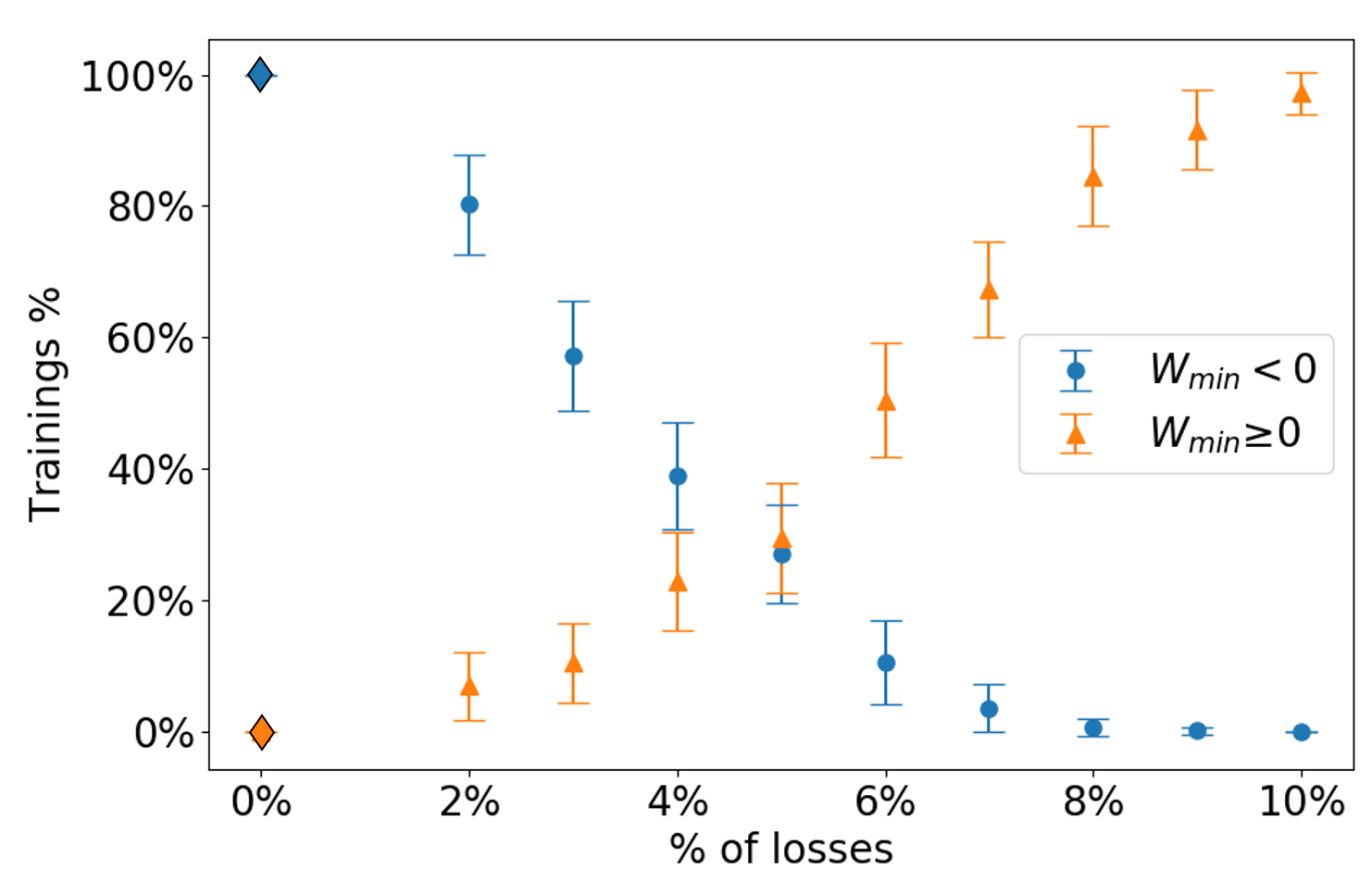}} 
\caption{Effect of extra losses. Top: $(2\pi)^2 W_{min}$ for different percentages of introduced losses as estimated by the MaxLik algorithm. Error bars are computed over $100$ repetitions. Bottom: Percentage of training sets identifying a negative state (blue circles) or a positive state (orange triangles) for at least $95$ out of $100$ inputs. Error bars are computed over $50$ runs of the whole process. In both panels diamonds corresponds to the experimental state, the circles (triangles) refer to states with $W_{min}<0$ ($W_{min}\ge0$).}
\label{fig3}
\end{center}
\end{figure}

We gathered evidence of such resilience by testing the consistency of the network as we introduce extra losses in the experimental data. We monitored the transition to a positive minimum value of the Wigner function as the losses increase. This can be simulated replacing a fraction of the quadrature data with data sampled from the vacuum state. We used the MaxLik method as a benchmark: it gives the results in Fig. \ref{fig3}, that shows $W_{min}$ as a function of the introduced losses. Each point represents the average over $100$ sets of 1000 quadratures, extracted at random from the same state. The error bar is the standard deviation on these replicas. As expected the Negativity decreases from the initial value, reaching positive values above $\sim 5 \%$.

We studied the same problem with the use of the NN algorithm. The training is operated generating the training data as before. Since we now underestimate the level of losses, the NN will not be optimized for this task, but we can rely on some robustness on its part. However, this property will strongly depend on the actual configuration reached with the training, hence on the random initial conditions. This implies that the test should be run over many instances of the training, so that one can assess the typical behaviour of the NN.

For each value of the loss, we carry out our analysis by feeding quadrature binned histograms derived from $100$ different repetitions of the lossy state to the NN  (the same one as for MaxLik benchmark). 
We then record the fraction of states $f_p$ with $W_{min}\ge 0$, and $f_n$ with $W_{min}< 0$ out of the $100$ attempts. To account for the variability of the training, we feed the same input data to $30$ differently trained NNs, and register the corresponding values of $f_p$ and $f_n$. In Fig.\ref{fig3}, we report the percentage of training instances for which $f_n\ge0.95$, and the same for $f_p\ge0.95$. These percentages do not sum to one, due to the presence of inconclusive results (see Supplementary Material \cite{SuppMat}). The NN method is thus able to identify reliably state with a negative Wigner function even when the actual amount of loss does not correspond exactly to that in the simulated data used for its training. Negativity is witnessed up to approximately $5\%$: remarkably, this is the threshold value at which the MaxLik shows the transition to a positive Wigner function.
 
In conclusion we have found that machine learning techniques can provide meaningful information on the Wigner Negativity even when limited data are available. 

The routine adoption of this method on large quantum cluster states is conditioned on the reliability of the data used for the training. 
We have shown that there exist specific instances in which the NN enjoys a given degree of flexibility.
The described method relies in fact on numerical simulations, since no encompassing analytical description is possible. Investigation of more generic states is the scope of future works, as the use of NN appears as a promising avenue for studying the behavior of large quantum states for which state tomography becomes impractical. In particular NN seem to be particularly useful to directly test specific quantum features of large multipartite systems without requiring the full reconstruction of the quantum states.   
This unleashes the potential of NN in quantum enabled technologies.
\nocite{*}

\begin{acknowledgments}
We thank I. Gianani for stimulating discussions. V.P. acknowledges financial support from the European Research Council under the Consolidator Grant COQCOoN (Grant No. 820079). N.T. acknowledges financial support of the Institut Universitaire de France.

\end{acknowledgments}

\bibliography{Bibliography1.bib}
\bibliographystyle{apsrev4-1}

\clearpage

\section{Supplementary Material}

\subsection{Simulated states for training NNs}

A crucial building block of our approach is the capability of simulating homodyne detection data for the class of quantum states under consideration. Here we explain in detail how the states, belonging to this class, are generated.

The first step in our simulation algorithm is to generate a $m$-mode Gaussian state, which is characterised by its mean field $\alpha \in \mathbb{R}^{2m}$ (describing a displacement in phase space) and its $2m \times 2m$ covariance matrix $V$. In this manuscript, we focus on the case where $\alpha = 0$, such that we can describe the full state by its covariance matrix. 

The covariance matrix can be decomposed through, first the Willamson decomposition, and then the Bloch-Messiah decomposition. As a result, we can write any arbitrary covariance matrix as 
\begin{equation}
    V= O_2 K O_1 \Delta O_1^t K O_2^t, 
\end{equation}
where $\Delta = {\rm diag}(\eta_1, \dots, \eta_m,\eta_1, \dots, \eta_m)$ is a diagonal matrix that describes the thermal noise, $K={\rm diag}(\sqrt{s_1}, \dots, \sqrt{s_m},1/\sqrt{s_1}, \dots, 1/\sqrt{s_m})$ is diagonal symplectic matrix that describes an inline squeezing operation, and $O_1$ and $O_2$ are symplectic orthonormal matrices that describe basis changes (i.e., passive linear optics operations).

In our simulations, $\Delta$, $K$, $O_1$, and $O_2$ are chosen randomly for each different state in the set of training data. For $\Delta$, we select $\eta_1, \dots, \eta_m$ from a uniform distribution between $0$ and some maximal value $\eta_{\rm max}$ (here we set $\eta_{\rm max} = 1.1$). A similar strategy holds for $K$, where $10\log(s_1), \dots, 10\log(s_m)$ are chosen randomly from a uniform distribution (i.e., the squeezing values are uniformly distributed on a ${\rm dB}$ scale), between zero and some maximal squeezing value $s_{\rm max}$ (here we set $s_{\max} = 8 {\rm dB}$). The $2m \times 2m$ basis changes $O_1$ and $O_2$ are generated via the Haar measure on the unitary $ m \times m$ matrices. To generate $O_j$, we sample a random $m \times m$ unitary $U_j = X_j + iY_j$, subsequently, we create a symplectic orthogonal matrix by setting
\begin{equation}
    O_j = \begin{pmatrix}X_j & Y_j \\ -Y_j & X_j\end{pmatrix}.
\end{equation}
In our specific algorithm, the second basis change $O_2$ is optional, which means that for each state we randomly choose whether or not $O_2$ is implemented. This choice is implemented to ensure that we also probe states in the basis where squeezing is ``local''. This is of particular interest for non-Gaussian states.

To approach realistic experimental conditions, we additional add the possibility to manually add losses (which may be known in a realistic experimental setting). This is done by controlling an additional parameter $\lambda$, and changing the covariance matrix according to
\begin{equation}
    V \mapsto (1-\lambda)V +\lambda \mathds{1}.
\end{equation}
This option is used in our analysis of experimental data, where we set $\lambda = 0.12$ in accordance with the experimental estimates.\\

Once the algorithm has generated the Gaussian state by sampling a covariance matrix $V$, a degaussifying operation is implemented two thirds of the times, choosing with equal probability among two different operations. As a non-Gaussian operation, we consider either photon addition or photon subtraction. When only a single photon is added or subtracted, the resulting Wigner function can be calculated analytically [49]. To implement either of these operations, we choose a random normalised vector $g \in \mathbb{R}^{2m}$, which represents the mode in which the photon is to be subtracted or added. Concretely, we then obtain the Wigner function (where ``$+$'' refers to photon addition and ``$-$'' to photon subtraction)
\begin{align}
W^{\pm}(\beta) 
%&= \frac{1}{(2\pi)^{2m}}\int_{\mathbb{R}^{2m}} {\rm d}^{2m}\alpha \, \chi(\alpha)e^{-i (\beta, \alpha)}\\
 &= \frac{1}{2} \Big[(\beta, V^{-1} A^{\pm}_{g} V^{-1} \beta) -  {\rm tr}(V^{-1}A^{\pm}_{g}) + 2 \Big]W_0(\beta) \label{eq:WignerFunction},
\end{align}
where $\beta \in \mathbb{R}^{2m}$ is a point in the optical phase space, and $(\beta,\alpha)$ denotes the scalar product. $W_0(\beta)=(2\pi)^{-m} (\det V)^{-1/2} \exp\left(-(\beta, V^{-1}\!\beta)/2 \right)$ is the initial Gaussian state's Wigner function.
\begin{equation}\label{eq:AMat}\begin{split}A^{\pm}_{g}= 2\frac{(V \pm \mathds{1})(P_{g} + P_{J\!g})(V \pm \mathds{1})}{{\rm tr}\{ (V \pm \mathds{1})(P_{g} + P_{J\!g})\}},\end{split}
\end{equation}
where $J$ denotes the symplectic form that fixes the structure of phases space, $P_g$ is a projector on the amplitude quadrature of mode $g$, and $P_{Jg}$ projects on its phase quadrature.\\

Once the state is degaussified (or not), we can directly extract its minimal value form (\ref{eq:WignerFunction}) and check whether or not it is negative. Furthermore, we can then calculate the marginals of the Wigner function via the method described in [49]. For any choice of homodyne detector phases, this allows us to deduce an $m$-dimensional probability distribution that describes the correlated outcomes of that specific measurement configuration. A basic rejection sampling algorithm can then generate the required number of joint quadrature measurements, and add them (together with the associated label for Wigner-negativity) to the set of training for the NN.

\subsection{Details on the NN algorithm}

\subsubsection{Input data}

\begin{figure*}[ht!]
\begin{center}
{\includegraphics[width=\textwidth]{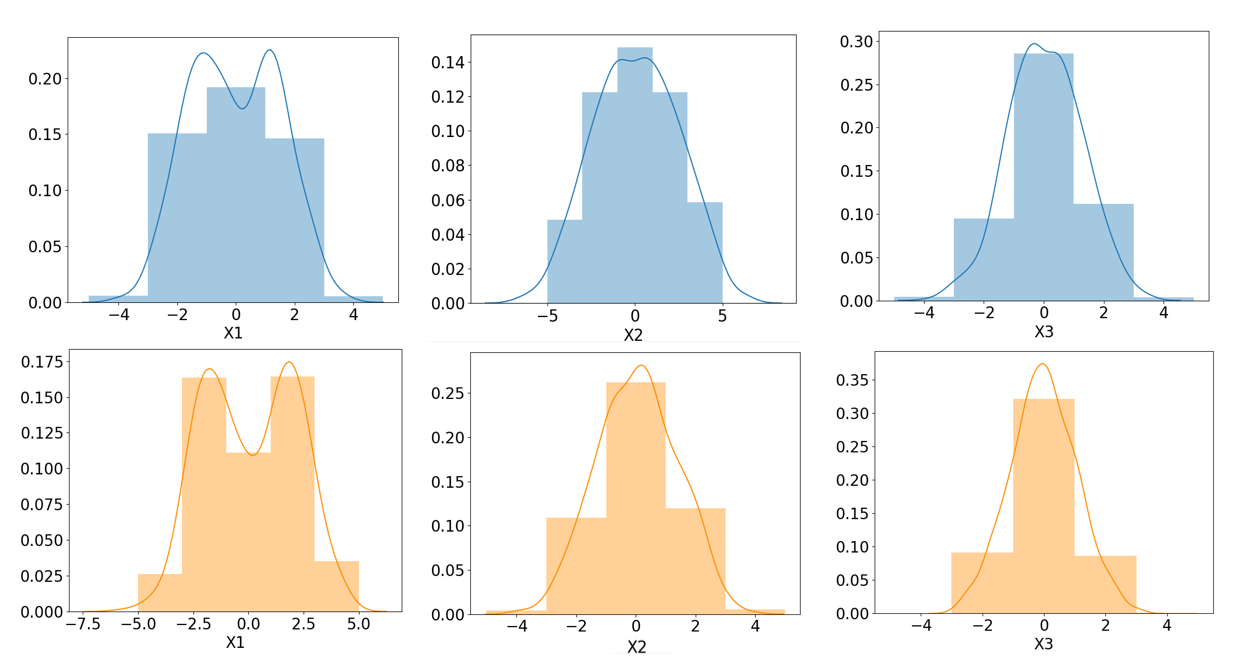}}
\caption{Binning of the quadrature histograms. The panels show the analytical form of the marginal distributions for the quadratures $X_1, X_2, X_3$ of the $m=3$ modes state. The top panel correspond to a state with $W_0=1$ and the bottom one to a state with $W_0=0$.}
\label{fig_s1}
\end{center}
\end{figure*}

\begin{figure}[h!]
{\includegraphics[width=\columnwidth]{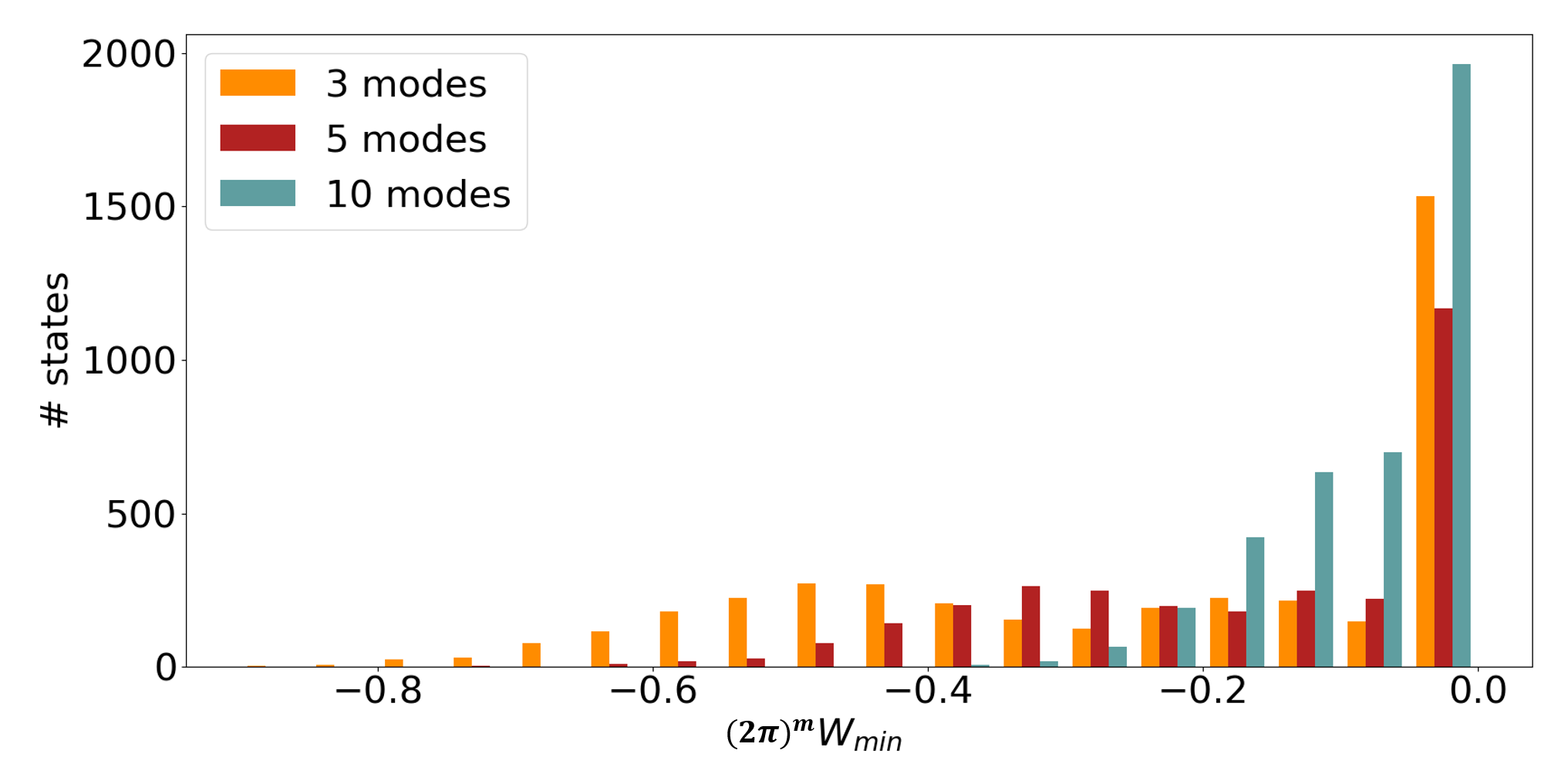}}
\caption{Number of states in the training set with a value of $(2\pi)^mW_{min}$ in the interval indicated on the $x$ axis. The three histograms show the values obtained for the three different configurations studied.}
\label{fig_s2}
\end{figure}

The input layer of our NN accepts the histograms of the quadrature measurements, such as those in Fig.\ref{fig_s1} that refers to one particular phase value. The quadratures are ranged in $5$ bins, with values between $-5$ and $5$. For each state in the training set the occurrence frequencies of such quadrature histograms are fed to the NN and, in order to associate to each training example the corresponding label $W_0$, the minimum value of the overall multimode Wigner function is computed.
We ensure that in the training set the number of states with and without Wigner-negativity are represented with almost equal weights. 
A clear indication of this can be obtained looking at the distribution of the rescaled Wigner minima of all the $4000$ states in the training set. Considering the three configurations studied, corresponding to states with a number of modes $m = 3, 5$ and $10$, the number of states with a value of $(2\pi)^mW_{min}$ in the range indicated on the $x$ axis are reported in Fig. \ref{fig_s2}.

A further improvement on the network performances has been obtained inserting a cut-off on the maximal negativity of the state in order to be considerate negative. The value of the cut-off has to be adapted depending on the number of modes $m$ of the states considered since the maximal negativity decreases as $m$ increases (as Fig. \ref{fig_s2} shows). Removing the states with a maximal negativity between $-0.1/(2\pi)^m$ and $0$ the network accuracy on the final estimation increases by a factor of $2-3\%$.

\subsubsection{Optimization step}

We implemented our algorithm using the python libraries for neural networks \emph{keras} and \emph{tensorflow}.

The vast majority of machine learning based minimization algorithms make use of gradient descent, an optimization algorithm that uses the first order derivative of the loss function to obtain its minimum. In our algorithm to minimize the loss function we use an optimizer inspired but different from the simple gradient descent, obtaining a big advantage in terms of the convergence time in the training step.  The minimization of the loss function is achieved by virtue of ``Adam'' optimizer [55]. This algorithm has been defined as a combination between stochastic gradient descent (SGD) with momentum, that randomly picks one data point from the whole data set at each iteration to reduce the computations enormously, and root mean square propagation (RMSprop) since it uses the squared gradients to scale the learning rate. Adam calculates the exponentially moving averages of the gradient evaluated on a selected mini-batch, adapting the parameters $\{\theta\}$ in the equations:

\begin{equation}
\begin{split}
m_t&=\beta_1m_{t-1}+(1-\beta_1)\nabla_\theta L_t(\theta_{t-1})\\
v_t&=\beta_2v_{t-1}+(1-\beta_2)\nabla_\theta L^2_t(\theta_{t-1}),
\end{split}
\end{equation}
where $\nabla_\theta L_t(\theta_{t-1})$ is the gradient on a mini-batch of the loss function, $\beta_1=0.9$ and $\beta_2=0.999$ are fixed to these values to obtain its best performance. The algorithm then implements a bias correction in order to recover the right estimation values:
\begin{equation}
\begin{split}
\hat{m}_t&=\frac{m_t}{1-\beta^t_1}\\
\hat{v}_t&=\frac{v_t}{1-\beta^t_2}.
\end{split}
\end{equation}
The parameters are then updated in the following way:

\begin{equation}
\theta_t=\theta_{t-1}-\eta\frac{\hat{m}_t}{\sqrt{\hat{v}_t}+\epsilon}
\end{equation}
where $\eta$ is the learning rate.

\subsubsection{Further performance tests}

As customary for classification algorithms we adopt two performance tests, both aiming at identifying the presence of false positives and negatives in the classification. This means that a state with $W_{min}<0$, correctly identified as $W_0=1$ is a true positive for the test. Instead, a state with $W_{min}\ge0$, misidentified as $W_0=1$ is a false positive. Therefore, the relevant quantities are the numbers of true positives ($T_p$), of true negatives ($T_n$), of false positives ($F_p$), and false negatives ($F_n$). These are used to define the recall $r=\frac{T_p}{T_p+F_n}$ and the specificity $s=\frac{T_n}{T_n+F_p}$.
 
\begin{figure}[h!]
\begin{center}
{\includegraphics[width=\columnwidth]{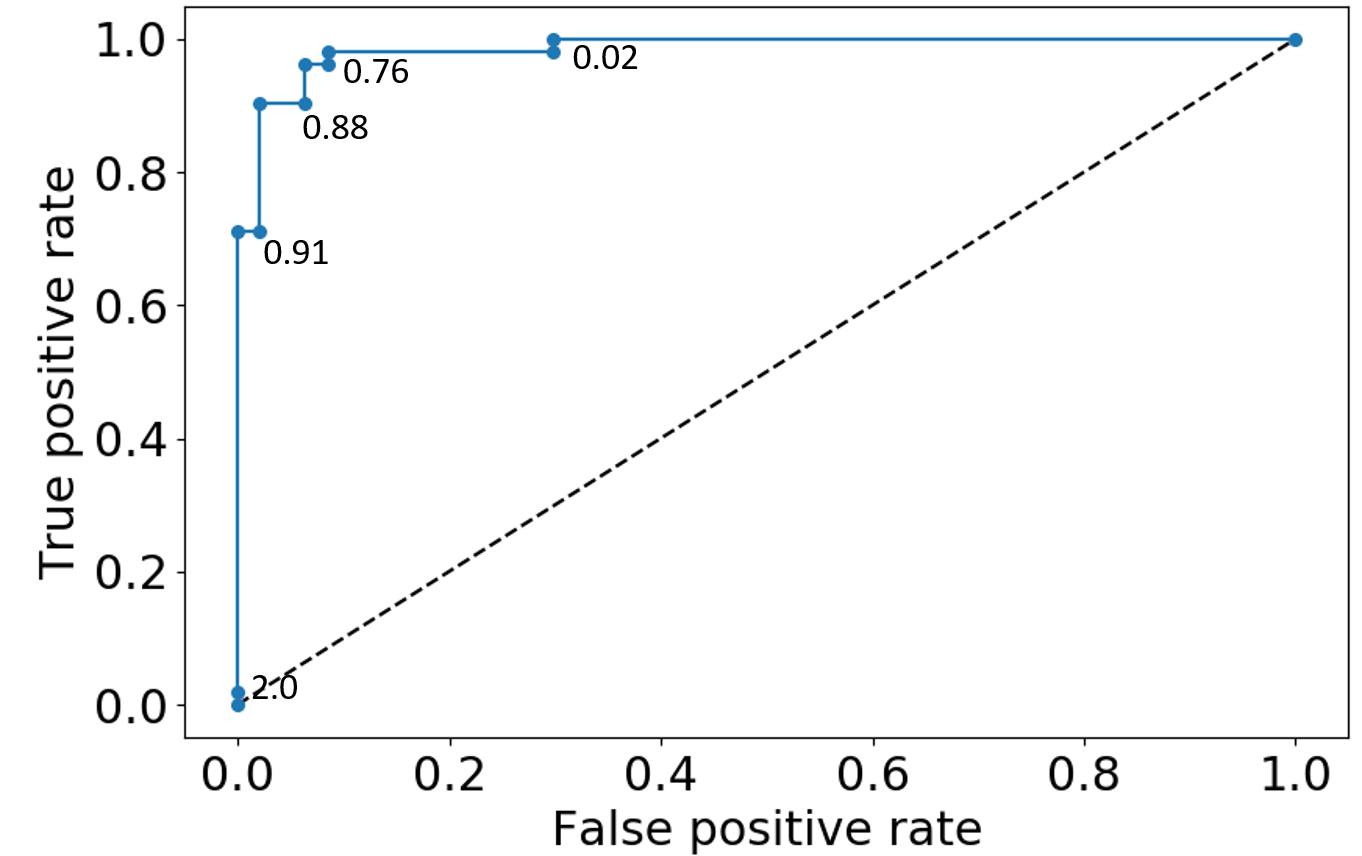}}
\caption{Receiver operating characteristic curve showing the classification performances for different thresholds values. The different points show the reliability of the NN classification. The dashed line represents the behaviour of a no-skill classifier.}
\label{fig_roc}
\end{center}
\end{figure}

The first test concerns the Receiver operating characteristic (ROC) curve which plots the true positive rate $r$ as a function of the false positive rate $1-s$ for different thresholds values $P_{th}$.
The extreme cases give the dotted line for a random classifier, while the perfect classifier is represented by the single point of coordinates $(1-s=0,r=1)$.
The ROC curve for our model is shown in Fig.\ref{fig_roc} and illustrates the good quality of our classifier.  

\begin{figure}[h!]
\begin{center}
{\includegraphics[width=\columnwidth]{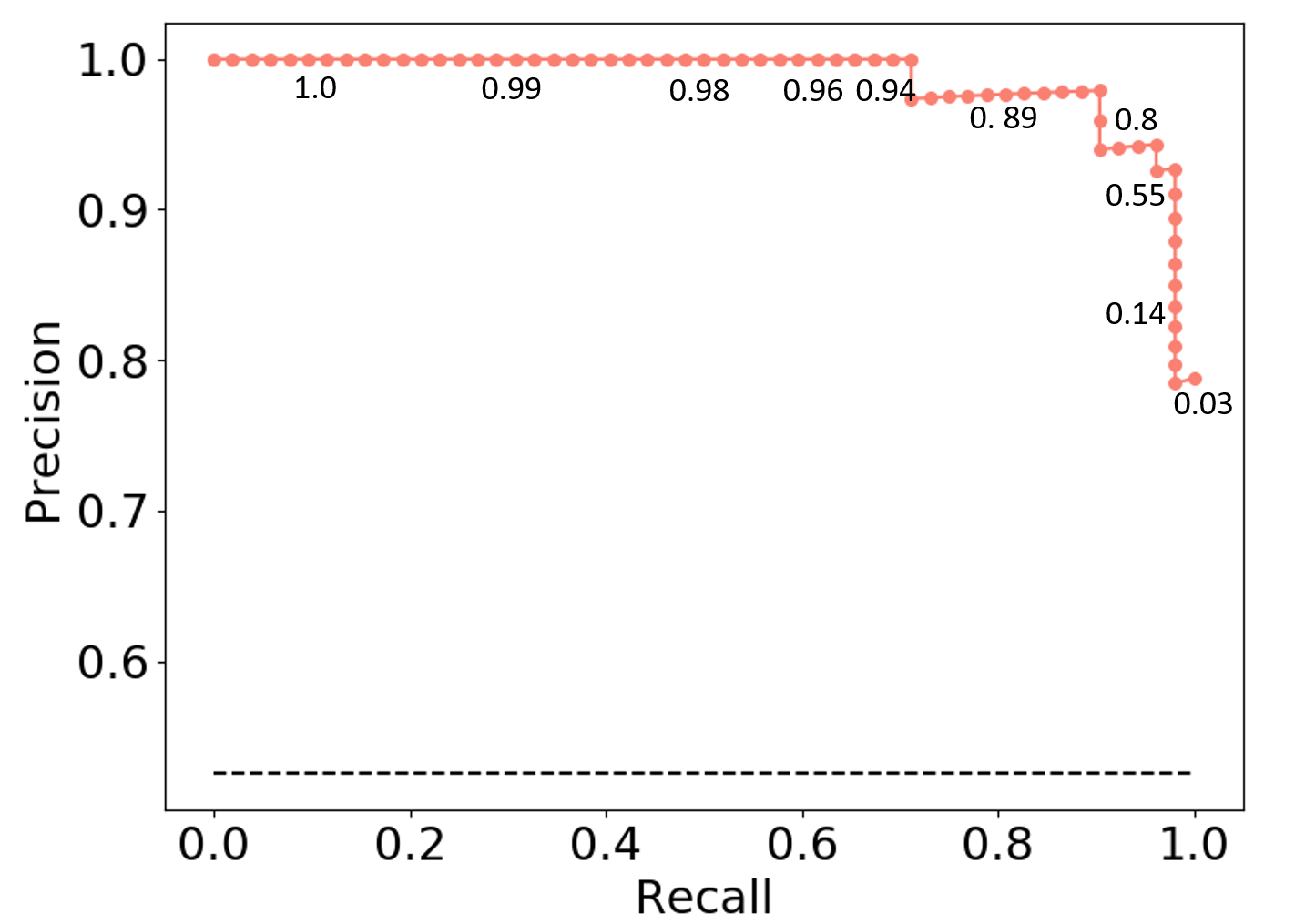}}
\caption{Precision vs. recall curve for different thresholds, it separates the analysis on the two classes identifying if the instances classified as true or negative are correctly classified.}
\label{fig_prec}
\end{center}
\end{figure}

The second test is particularly suited for training sets with an imbalance in between the two classes of states. In this scenario the accuracy could remain high even if the less represented class is wrongly or randomly identified. It is useful to study the precision $p=\frac{T_p}{T_p+F_p}$ as a function of the recall for varying $P_{th}$. The results are shown in Fig.\ref{fig_prec}, demonstrating that our curve is well above the random limit of $p=0.5$.  

\subsubsection{Histograms for extra losses}

\begin{figure}[h]
\begin{center}
{\includegraphics[width=\columnwidth]{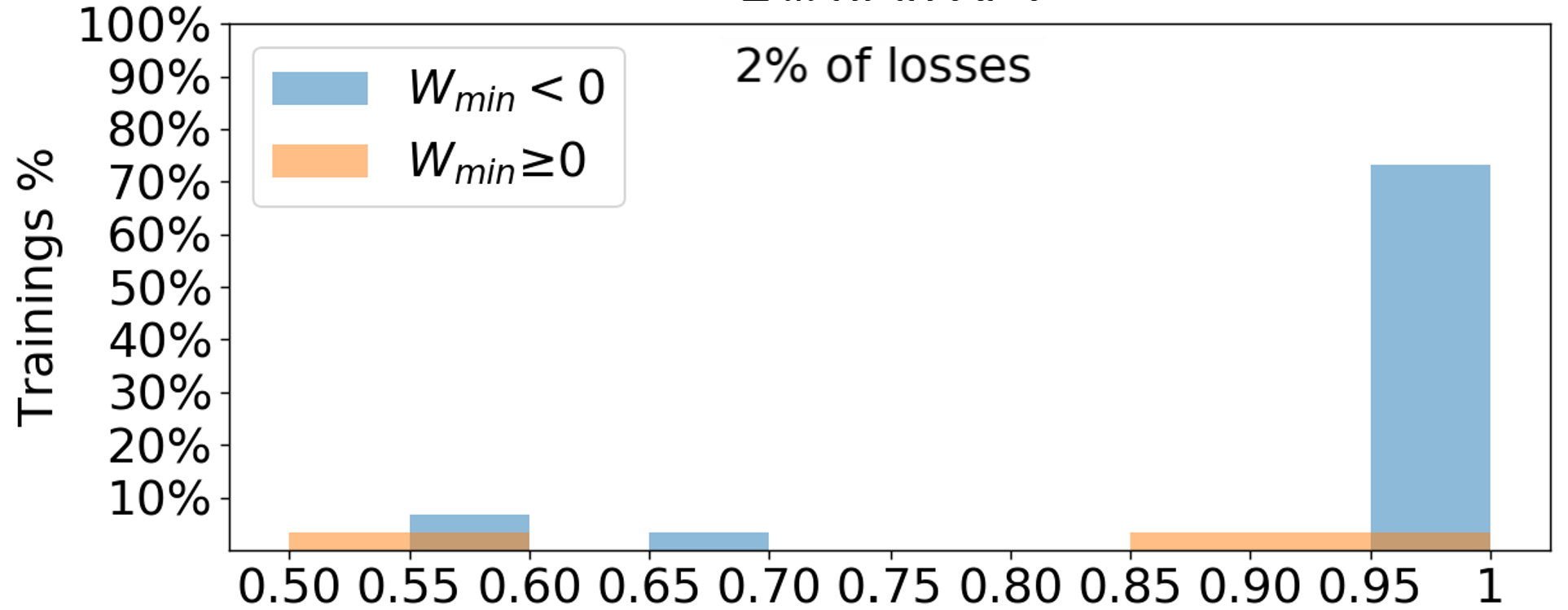}} \\
{\includegraphics[width=\columnwidth]{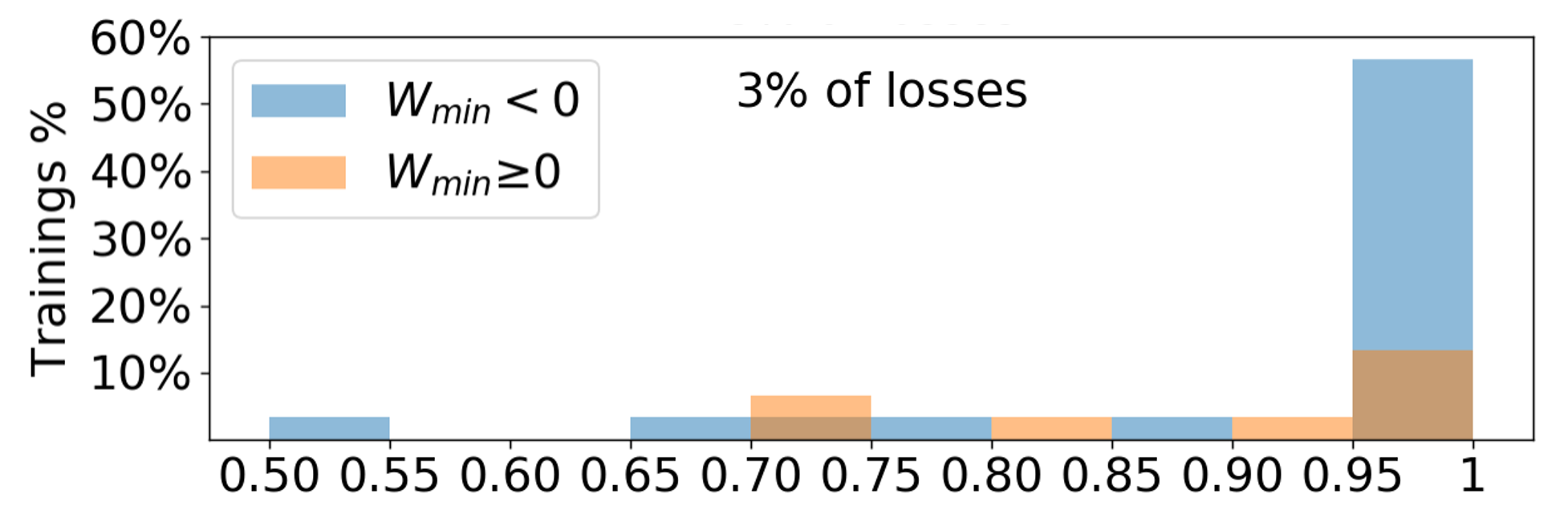}} \\
{\includegraphics[width=\columnwidth]{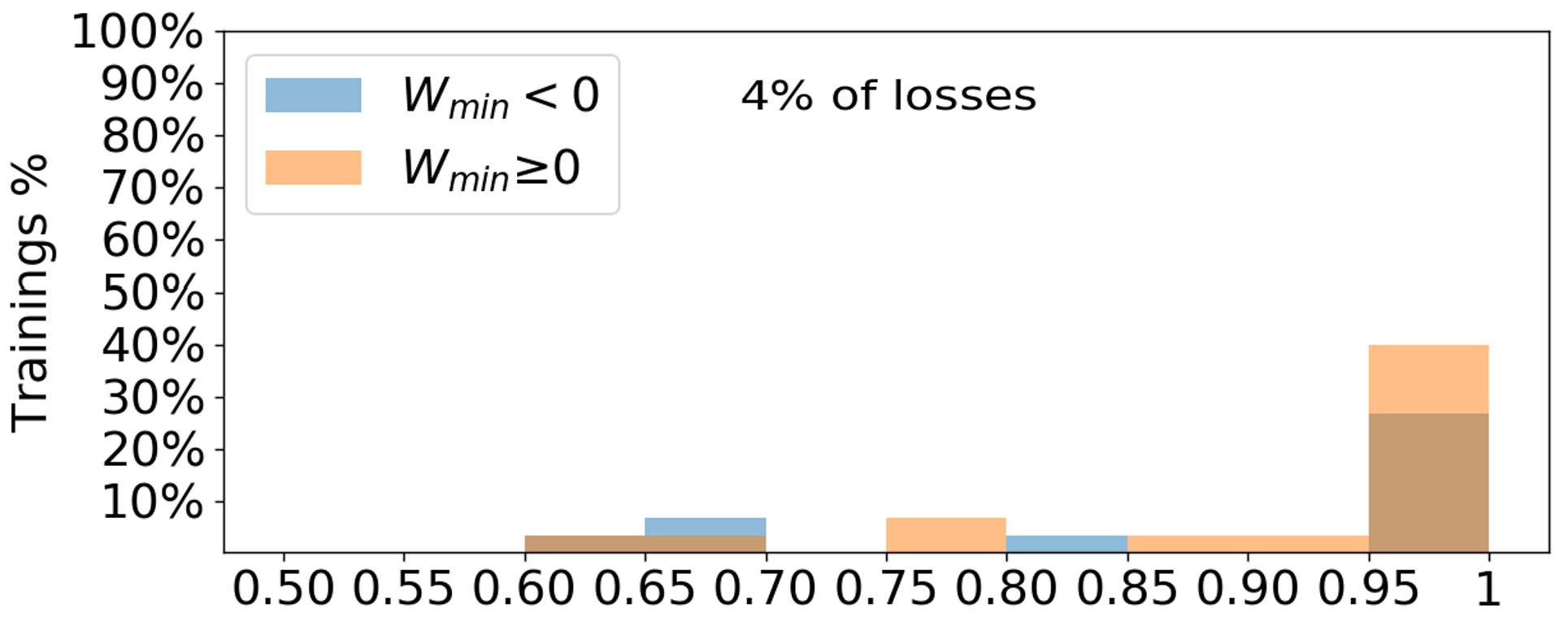}} \\
{\includegraphics[width=\columnwidth]{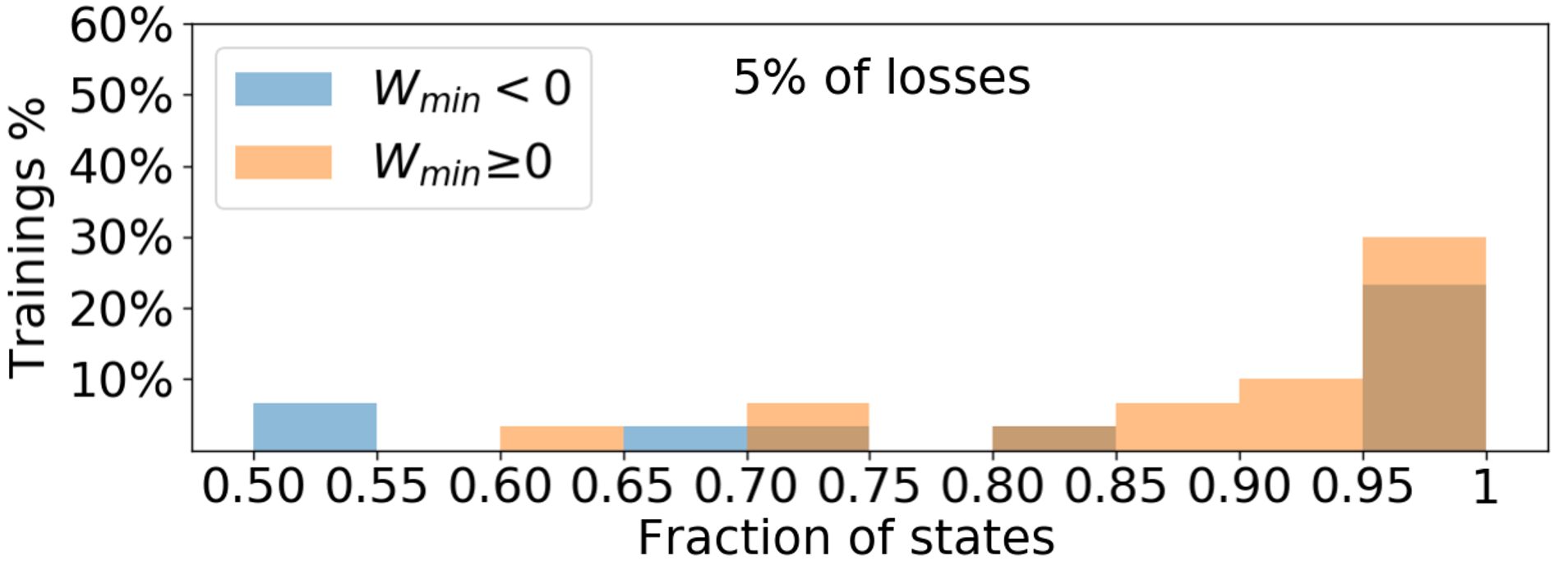}} \\
{\includegraphics[width=\columnwidth]{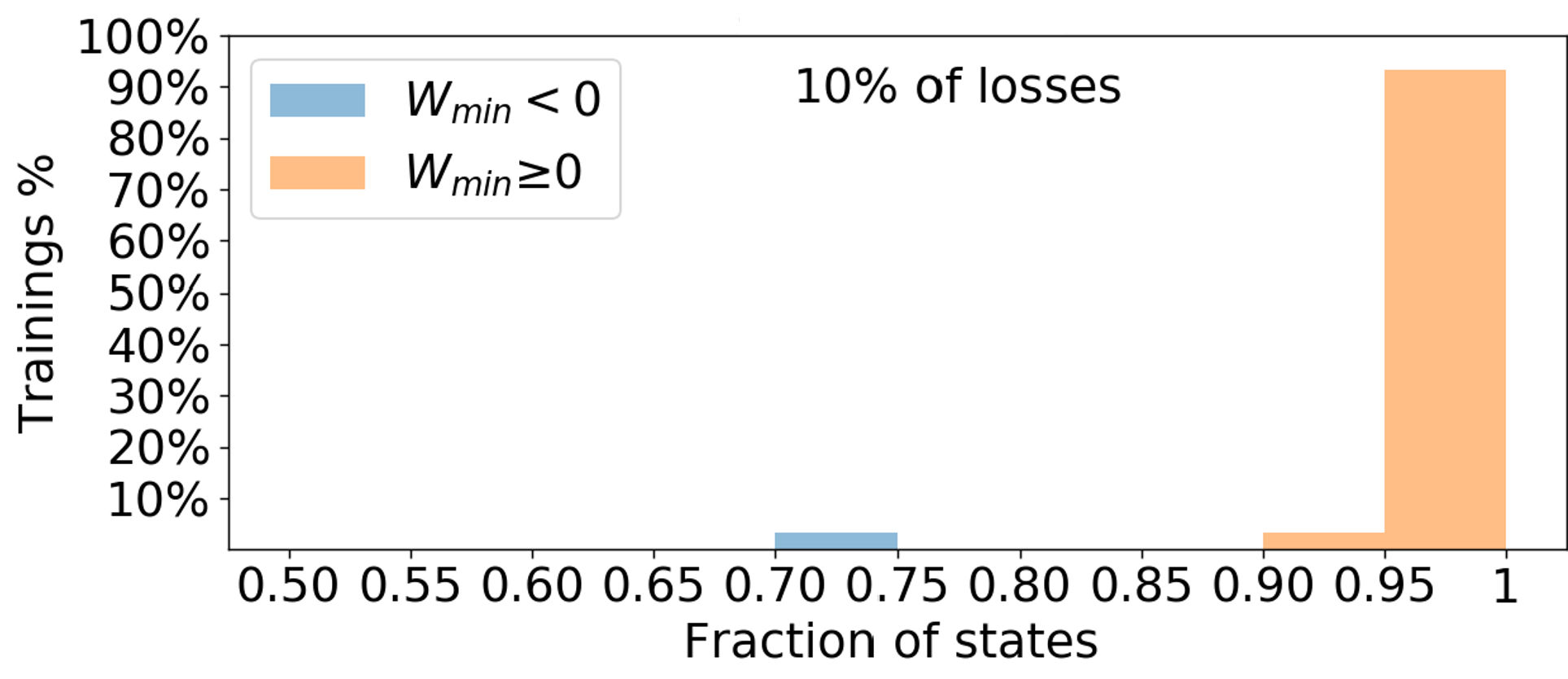}} 
\caption{Histograms of the classification of our experimental state for increasing values of added losses.}
\label{fig7}
\end{center}
\end{figure}

We investigate the NN behaviour in classifying states when extra losses in the quadrature data are introduced compared to the training ones. To this purpose we randomly replace the original experimental data with a fraction of quadrature data sampled from the vacuum state generating $100$ replicas of such lossy states. After the NN has been trained with states without the introduction of the extra losses generated as described in the main text, we look at its response when it is fed with the histograms coming from the replaced data for all the replicas. Due to this difference between the training data and the ones on which the network is tested, we also take into account the variability coming from $30$ different trainings of the algorithm studying the response's variability in evaluating these replicas coming from the same original state. In Fig.\ref{fig7} the full statistics of the different training is reported.
For each level of introduced extra losses, we register the percentage of trainings identifying a negative (positive) Wigner function, reported in blue (orange), on the fraction $f_n$ ($f_p$) of $100$ states indicated on the $x$ axis.

The data in Fig.3 in the main text are obtained from histograms such as those in Fig.\ref{fig7} reporting only the training percentages which obtained a fraction of positive $f_p$ and negative $f_n$ states exceeding $95\%$  

\subsection{Experimental Wigner function}
The experimental data used here are derived from the experience presented in [53]. The two-mode entangled state is generated in a parametric down conversion process in a resonant cavity which is synchronously pumped via the second-harmonic of a femto-second Ti:Sapphire laser (pulse duration of 90 fs, central wavelength of 795 nm, at a repetition rate of 76 MHz). The setup generates several squeezed time-frequency modes [4], we choose two of them and the appropriate basis change in order to generate a two-mode squeezed state (EPR state). The generated state undergoes a photon subtraction process via sum-frequency interaction in a second non-linear crystal where it interacts with a coherent gate-beam. The detection of one up-converted photon heralds single -photon subtraction from the mode which is defined by the mode of the gate beam [56],  and that we choose to be one of the two entangled modes.
The resulting state is well approximated by a separable state of a vacuum state in the first mode and a single photon state in the second mode. This can be explained by considering that the two-mode squeezed state is described in the photon-number basis as superposition of even photon-number in the two modes, with decreasing probability amplitude  for larger photons-number.  So when a single-photon subtraction is done on one of the modes, the presence of this photon that has been subtracted is strongly correlated with the presence of one photon (and only one) in the other mode.
The reconstructed Wigner function of each mode is shown in the last line of Fig. 2 b) of [53] along with the value of $2 \pi W(0,0)$. Here we use the same homodyne data of the two independently measured modes, to check the Wigner negativity of the two-mode state via the NN method and via a two-mode MaxLik reconstruction. In the latter we assume no correlation between the two modes. The reported minimal value of the Wigner function $(2 \pi)^2 W_{min}$ in the first point of the upper part of Fig.3 is in fact equivalent to the product of the two value of $2 \pi W(0,0)$ for the $EPR_0$  and $EPR_1$ modes in the last line of Fig. 2 b) of [53].

\end{document}